\documentclass[conference]{IEEEtran}

\usepackage{cite}
\usepackage{threeparttable}
\usepackage{amsmath,amssymb,amsfonts}
\usepackage{algorithm, algorithmic}
\usepackage{graphicx}
\usepackage{threeparttable}
\usepackage{textcomp}
\usepackage{balance}
\usepackage{xcolor}
\usepackage{url}
\usepackage{hyperref}
\usepackage{textgreek}
\usepackage{array}
\usepackage{makecell}
\newcolumntype{L}{>{\phantom{$\mathbin{-}$}$}l<{$}}

\newcolumntype{M}[1]{>{\centering\arraybackslash}m{#1}}
\newcolumntype{N}{@{}m{0pt}@{}}

\renewenvironment{thebibliography}[1]{
  \begin{oldthebibliography}{#1}
    \setlength{\itemsep}{0em}
    \setlength{\parskip}{0em}
}
{
  \end{oldthebibliography}
}
\def\BibTeX{{\rm B\kern-.05em{\sc i\kern-.025em b}\kern-.08em
    T\kern-.1667em\lower.7ex\hbox{E}\kern-.125emX}}
\usepackage{pifont}

\begin{document}

\title{Open Information Extraction: A Review of Baseline Techniques, Approaches, and Applications}

\author{\IEEEauthorblockN{Serafina~Kamp$^+$}
\IEEEauthorblockA{\textit{Department of EECS} \\
\textit{University of Michigan}\\
Ann Arbor, US \\
serafibk@umich.edu}
\and
\IEEEauthorblockN{Morteza~Fayazi$^+$}
\IEEEauthorblockA{\textit{Department of EECS} \\
\textit{University of Michigan}\\
Ann Arbor, US \\
fayazi@umich.edu}
\and
\IEEEauthorblockN{Zineb~Benameur-El}
\IEEEauthorblockA{\textit{Department of EECS} \\
\textit{University of Michigan}\\
Ann Arbor, US \\
zinebbe@umich.edu}
\and
\IEEEauthorblockN{Shuyan~Yu}
\IEEEauthorblockA{\textit{Department of EECS} \\
\textit{University of Michigan}\\
Ann Arbor, US \\
shuyan@umich.edu}
\and
\IEEEauthorblockN{Ronald~Dreslinski}
\IEEEauthorblockA{\textit{Department of EECS} \\
\textit{University of Michigan}\\
Ann Arbor, US \\
rdreslin@umich.edu}
}
\maketitle
\def\thefootnote{+}\footnotetext{The contributions of the first two authors are equal.}

\begin{abstract}
With the abundant amount of available online and offline text data, there arises a crucial need to extract the relation between phrases and summarize the main content of each document in a few words. For this purpose, there have been many studies recently in Open Information Extraction (OIE). OIE improves upon relation extraction techniques by analyzing relations across different domains and avoids requiring hand-labeling pre-specified relations in sentences. This paper surveys recent approaches of OIE and its applications on Knowledge Graph (KG), text summarization, and Question Answering (QA). Moreover, the paper describes OIE basis methods in relation extraction. It briefly discusses the main approaches and the pros and cons of each method. Finally, it gives an overview about challenges, open issues, and future work opportunities for OIE, relation extraction, and OIE applications.
\end{abstract}

\maketitle

\section{Introduction}
As huge amounts of offline and online text documents (e.g. digital libraries) are available, there have been a lot of efforts to better organize and extract this information~\cite{witten2005kea, SentimentClassificationReview}. Finding the relations between text entities leads to many applications such as summarizing the main content of documents and automatic answer retrieval from unstructured text. For this purpose, relation extraction methods extract the relation between the two entities within one sentence in a text.

Relation extraction has broad applications, including Knowledge Base (KB) population for question answering systems, bio-text mining, and information extraction from clinical texts~\cite{rel_deep_survey,rel_sup_deep_gcn_protein,mehryary2018potent,rel_sup_clinical_extract,rel_sup_clinical_outcome}. Despite supervised learning being the common approach to solve information extraction problems, it is heavily dependent on manually annotated training data. In response, Open Information Extraction (OIE) refrains from using relation training data which is not bound by a fixed relation vocabulary. Its main principle is not only to extract arguments, but also to extract relation phrases from the text.

Even though relation extraction methods are mainly limited to recognizing relations from single input sentences~\cite{bach2007review}, the advancement of OIE relies on them. For example, relation extraction techniques can be leveraged to populate a KB which is used in semi-supervised OIE methods. OIE methods then learn to find important and frequent relations in a document in contrast to only recognizing a single relation in one sentence. An important part of using large KBs is canonicalization which helps in structuring and accessing data within such KBs~\cite{vashishth2018cesi}.

With the recent progress of OIE, several OIE-based applications have been developed, including knowledge base construction for Knowledge Graphs (KGs)~\cite{martinez2018openie} and minimizing redundant information for text summarization~\cite{christensen2013towards10}. OIE is an important preprocessing stage here because OIE systems extract simple relational facts that are necessary for these semantic task applications~\cite{mausam2016open, stanovsky2015openintermediate}.

The rest of the paper is organized as follows. Section \ref{sec:relation_extraction} talks about the OIE's baseline approaches in relation extraction. Section \ref{sec:open_information_extraction} discusses the main recent open information extraction and canonicalizing open knowledge base techniques. Next, Section \ref{sec:application} describes the semantic downstream applications of OIE that can be improved by incorporating OIE as intermediate steps. In each of these sections, the main challenges are discussed and some ideas about future works are suggested. Finally, the paper is concluded in Section \ref{sec:conclusion}.

\section{Relation Extraction}
\label{sec:relation_extraction}
Relation Extraction is the task of extracting a tuple, $t=$~$ (e_1,r,e_2)$, where $e_1,e_2$ are entities, the subject or object of the relation identified in a text, and $r$ defines the relation between these two entities~\cite{bach2007review}. The relation extraction methods can be categorized into supervised and semi-supervised learning techniques. These techniques look for relations in a text, based on labeled training examples. The influence of the training data limits the generalization capabilities of these approaches and motivates people to use OIE methods where more diverse relations can be discovered and extracted. OIE analyzes information in the whole of a document, while relation extraction usually focuses on relations found within one sentence. This means OIE has more generalization capability than relation extraction.

\subsection{General Terminology}

\subsubsection{Model Inputs} 
\begin{figure}[t]
\includegraphics[width=\columnwidth]{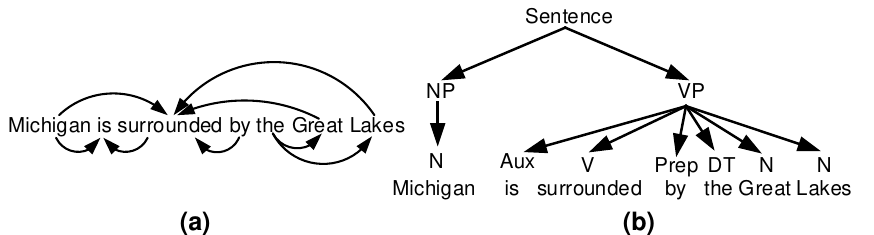}
\caption{Examples of a sentence in (a) dependency parse tree~\protect\cite{rel_sup_dep_kernel_short} and (b) full parse tree representation~\protect\cite{rel_sup_features}.}
\label{fig:rel_sup_tree}
\end{figure}

Sentences can be represented in various ways for different models. To this end, several terms are commonly used to describe these different types of representation. Part-of-Speech (POS) denotes the type of each word in a sentence, e.g. noun, verb, adjective, etc. POS is used in parse trees, a type of graph used to capture the syntactic information of a sentence.  The nodes of parse trees are words in the sentence that are labeled with their POS. This representation can be modified into a dependency parse tree which includes only directed connections between nodes to indicate the dependencies between them. An example of these two trees is shown in Figure~\ref{fig:rel_sup_tree}. Node labels can be generalized as word classes~\cite{rel_sup_dep_kernel_short} which are different ways to classify words in a sentence such as POS and hypernyms. In some neural network models, a numerical representation of sentences is needed, so word embeddings are used. This is the process of converting a sentence to a meaningful representation in a numerical vector format.

\subsubsection{Model Design Choices}
Different model architectures for relation extraction have some common design choices. Mikolov~\textit{et~al.}~\cite{rel_sup_word2vec} propose the skip-gram model which predicts the context of a given input word in a sentence. The model is trained with many different sentences to learn the context of different words. Then, when the model is given an input word, it outputs words that are typically in the context, surrounding the input word. The skip-gram model is used to train word2vec~\cite{rel_sup_word2vec}, which is a model that creates word embeddings for sentences. A better sentence representation can be gained by adding subsequent attention layers to the word embedding layer. An attention layer is a neural network layer that is used to learn which areas of the input are useful in understanding other areas of the input. So, in the context of sentence understanding, this would equate to understanding which words earlier in the sentence help one understand the current word being read.

\subsection{Supervised Learning}
In supervised methods, the relations are learned by training on many labeled samples of the desired relations. These methods can be divided into kernel-based, feature-based, and deep learning approaches. Kernel-based methods use a similarity function~\cite{talburt2015entity} to find related inputs. On the other hand, feature-based methods use grammatical structure and word patterns to find relations. A key difference between these two methods is the fact that kernel-based methods do not require defining which features show similarities between two inputs. In contrast, feature-based methods rely on explicitly defining features to measure the similarity of inputs~\cite{rel_sup_many_kernels,rel_sup_features}. Kernel- and feature-based methods are not popular in recent advances and they have been studied thoroughly in~\cite{bach2007review}. For this reason, we will first discuss the limitations of these methods and then review the recent state-of-the-art deep learning methods for supervised relation extraction.

\subsubsection{Kernel- and Feature-Based Approaches}
Typically, the input to both methods is a parse tree which stores relevant information, e.g. Part-of-Speech (POS) and the dependency of one word on another, about words in a sentence. The creation of this parse tree happens in a preprocessing stage with third-party tools, which could be prone to errors. An example of dependency of one word on another would be in the sentence: ``The sun is bright", where ``bright" modifies ``sun" such that ``sun" is dependent on ``bright". Kernel-based methods use a similarity function to calculate how related two inputs are~\cite{rel_sup_many_kernels}. Bunescu~\textit{et~al.}~\cite{rel_sup_dep_kernel_short} calculate the similarity between two relation dependency parse trees based on the number of common word classes (POS, dependency direction, etc.) for each component in the dependency parse tree. Feature-based methods rely on explicitly defined features to classify relation instances~\cite{rel_sup_features}. Tongtep~\textit{et~al.}~\cite{rel_sup_feature_thai} use three features
to classify different types of relations, including token space, number of named entities, and entity type. In feature-based models, features must be explicitly known which makes it difficult for such models to generalize and recognize new types of relations.

A more thorough review of these methods can be found in~\cite{bach2007review}. In general, both kernel- and feature-based methods suffer from not being able to generalize well to new or more complicated relations because of the reliance on defining a feature space and the preprocessing stage.

\subsubsection{Deep Learning Approaches}
Since choosing relevant features for a relation extraction model is ambiguous, recent works have explored supervised deep learning where features are inherently implicitly chosen. These approaches can be treated as relation classification tasks where the training data is a set of input sentences and the label is the relation the sentence conveys~\cite{rel_sup_CNN, rel_deep_survey}. An important aspect of these deep learning models is how an input sentence is formatted. A common input formatting approach is word embedding~\cite{rel_deep_survey} which can be created from the word2vec~\cite{rel_sup_word2vec} or the Bidirectional Encoder Representations from Transformers (BERT) models~\cite{rel_sup_deep_bert}.

Mikolov~\textit{et~al.}~\cite{rel_sup_word2vec} introduce word2vec for the task of creating a static, learned embedding of words. Word2vec uses a skip-gram model to learn ``nearby" words for a given input word expressed in a numerical vector representation. In other words, the model learns to predict surrounding context words based on an input word. This training setup results in learning individual vector representations of each word which preserves the relations found in typical contexts of that word. For instance, the vector representation of ``Berlin" is geometrically nearby the summation of the vector representation of ``Germany" and the vector representation of ``capital"~\cite{rel_sup_word2vec}. This is a natural baseline for relation extraction tasks as entities that are related are likely to be nearby each other in this numerical vector formalization. Nguyen and Grishman~\cite{rel_sup_CNN} utilize this word embedding along with a Convolutional Neural Network (CNN) \cite{albawi2017understanding,fayazi2021applications}. They conclude that using a max pooling layer in CNNs helps emphasize the important features of the input sentence embedding. In comparison with previous methods that use explicitly defined features~\cite{rel_sup_feature_thai,rel_sup_features}, this CNN implicitly gathers and finds the most important features from the word2vec embedding and improves feature selection for the relation extraction task.

Devlin~\textit{et~al.}~propose the BERT model~\cite{rel_sup_deep_bert}, which is used for pre-training in some deep relation extraction models. In this context, pre-training consists of mapping input sentences to a meaningful embedding. BERT uses a Masked Language Model (MLM) approach to capture the left and right side context of each word in a sentence during training. BERT is also trained with a next-sentence prediction stage. The MLM randomly chooses a subset of input tokens to hide during training while the next sentence prediction stage randomly switches the order of a subset of sentences. BERT is a language transformer and, typically, transformers are unidirectional. So, BERT is a significant advancement because relation instances often rely on information from both side contexts of entities. These training steps help the model learn the surrounding contexts of entities which is powerful in relation extraction~\cite{rel_sup_mtb,rel_sup_sota_semEval}.

Tao~\textit{et~al.}~and Baldini Soares~\textit{et~al.}~\cite{rel_sup_sota_semEval,rel_sup_mtb} enhance the BERT foundation with additional training circumstances. Tao~\textit{et~al.}~\cite{rel_sup_sota_semEval} define ``syntactic indicators" as an input to their model while Baldini Soares~\textit{et~al.}~\cite{rel_sup_mtb} introduce Matching The Blank (MTB) training. Syntactic indicators are defined as the words or phrases that provide semantic information about the two entities of interest~\cite{rel_sup_sota_semEval}. They find that the addition of syntactic indicators to the BERT input leads to a consistent increase in F-1 Score. In MTB, the input sentence is augmented by replacing a random subset of entities with ``[BLANK]". This follows the intuition that a relation pair occurs frequently in a large corpus of text in slightly different contexts. Learning which words have been replaced by ``[BLANK]" can provide rich semantic details about the relations. This leads the model to learn how to generalize relations without a large amount of human-labeled relation pairs. As such, MTB can be classified as more of a semi-supervised approach, but Soares~\textit{et~al.}~\cite{rel_sup_mtb} compared this method to their own completely supervised method where no ``[BLANK]" tokens added to the inputs. This shows that it is possible to achieve similar performance to a completely supervised model without needing many labeled training samples.

\subsection{Semi-Supervised}
\begin{figure}[t]
\includegraphics[width=\columnwidth]{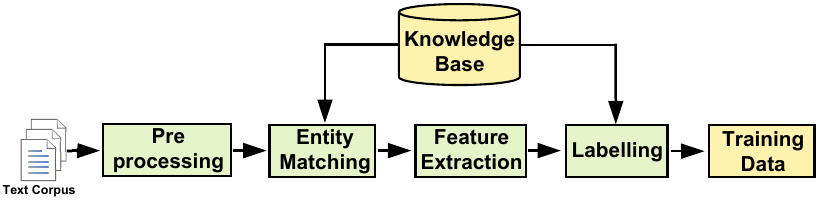}
\caption{The pipeline of creating training data in semi-supervised relation extraction~\protect\cite{rel_semi_review}.}
\label{fig:rel_semisup}
\end{figure}

Semi-supervised learning follows the idea of using less manually labeled training data. Models start with a small set of manually defined rules which are used to build more patterns during training~\cite{bach2007review}, it is shown in Figure~\ref{fig:rel_semisup}. Early successful works, such as Snowball~\cite{rel_semisup_snowball}, have been reviewed thoroughly in~\cite{rel_semi_review}, so we will focus on recent works and deep learning models. We first review non deep learning semi-supervised methods and then evaluate the advantages gained from deep learning semi-supervised models.

\begin {table*}[t]
\scriptsize
\begin{center}
\begin{threeparttable}
\caption{Relation extraction methods summary. \tnote{$\dagger$}}
\centering
\def\arraystretch{1.3}\tabcolsep 2pt
\label{table:relation_extraction}
\begin{tabular}{|M{8mm}|M{20mm}|M{65mm}|M{45mm}|M{30mm}|}
\hline
\textbf{Work} & \textbf{Supervision Level} & \textbf{Main Technique} & \textbf{Scope} & \textbf{Dataset}\\ 
\hline
\cite{rel_sup_dep_kernel_short} & Supervised & Shortest path between entities on a dependency parse tree & Relations where all the information about them is contained in the given sentence & ACE 2004~\cite{mitchell2005ace}\\ \hline
\cite{rel_sup_features} & Supervised & Analyzing different input representation of sentence inputs: Dependency, sequential, and syntactic parse trees. Sequential+syntactic parse tree gives the best feature space & N/A & ACE 2004~\cite{mitchell2005ace}\\ \hline
\cite{rel_sup_feature_thai} & Supervised & Using features of token, entity count, and entity type to classify relation patterns & Works well for person-action, action-person, action-location, and location-action relations & Online Thai news Sources \\ \hline
\cite{rel_sup_clinical_extract} & Supervised & NLP and pattern matching & Clinical notes & Randomly selected patient notes~\cite{rel_data_extract}\\ \hline
\cite{rel_sup_CNN} & Supervised & CNN with word2vec input & N/A & SemEval 2010~\cite{hendrickx2019semeval} ACE 2005~\cite{walker2006ace}\\ \hline
\cite{rel_sup_deep_gcn_protein} & Supervised & GCN & Chemical-protein interactions & ChemProt~\cite{rel_data_chemprot}\\ \hline
\cite{mehryary2018potent} & Supervised & LSTM network & Chemical-protein interactions & ChemProt~\cite{rel_data_chemprot}\\ \hline
\cite{rel_sup_sota_semEval} & Supervised & Training: Syntactic indicators, Inference: Neural network with BERT input & N/A & SemEval 2010~\cite{hendrickx2019semeval}\\ \hline 
\cite{rel_sup_mtb} & Supervised & Training: MTB, Inference: Neural network with BERT input & N/A & FewRel~\cite{rel_data_fewrel}\\ \hline 
\cite{rel_semisup_snowball} & Semi-Supervised & Pattern generation from seed rules & Works well when training data is a small set of relation patterns & North American news text corpus~\cite{rel_data_news} \\ \hline
\cite{rel_semisup_luchs} & Semi-Supervised & Conditional random field & Hierarchical approach allows for efficient scanning of Wikipedia pages & Wikipedia\\ \hline
\cite{rel_semisup_minsup} & Semi-Supervised  & Knowledge base for numerical relations & Relations with one numerical entity & TACKBP 2014~\cite{rel_data_tackbp}\\ \hline
\cite{rel_semisup_pcnn} & Semi-Supervised & PCNN with word2vec input & N/A&Riedel 2010~\cite{rel_data_riedel}  \\ \hline
\cite{rel_semisup_ea} & Semi-Supervised & RNN variation and PCNN with entity and word attention layers & N/A & Riedel 2010~\cite{rel_data_riedel},  GDS~\cite{rel_semisup_ea} \\ \hline
\end{tabular}
\begin{tablenotes}\footnotesize
\item[$\dagger$] Abbreviation list: NLP: Natural Language Processing, CNN: Convolutional Neural Network, GCN: Graph Convolutional Network, LSTM: Long-Short Term Memory, BERT: Bidirectional Encoder Representations from Transformers, MTB: Matching The Blank, PCNN: Piecewise Convolutional Neural Network, RNN: Recurrent Neural Network.
\end{tablenotes}
\end{threeparttable}
\end{center}
\end{table*}

Hoffmann~\textit{et~al.}~\cite{rel_semisup_luchs} propose LUCHS which uses a Conditional Random Field (CRF)~\cite{rel_semisup_crf} to extract relation facts from Wikipedia. The model has a hierarchical approach by running a classifier to determine whether an input text is likely to contain a certain relation type, such as birthplace~\cite{rel_semisup_luchs}. It then searches the input text for instances of that relation type. These instances are used to expand the training set by providing new representations of the previously observed samples of that relation. This allows the model to scale and learn many more relation types than using manually labeled training data. Madaan~\textit{et~al.}~\cite{rel_semisup_minsup} propose and compare a supervised and a semi-supervised method, NumberRule and NumberTron, respectively. These models are focused specifically on numerical relation extraction where one entity of interest is a numerical value. NumberRule represents the inputs as the shortest dependency path from a parse tree as seen in kernel-based methods. Then, it checks for predefined relation keywords to extract a relation between given entities of interest. On the other hand, NumberTron uses a knowledge base in a semi-supervised approach to match the units of the target relation to entries with that unit in the KB. This increases the flexibility of the model as it does not solely rely on predefined keywords to find relation instances. Indeed, NumberTron outperforms NumberRule in their experiments and confirms the advantage of semi-supervised learning. The main limitation of this method is that it cannot extract the time scope of the numeric attribute. This limitation is important since numeric values associated with entities can change often and knowing how long the extracted relation is valid is necessary.

Jat~\textit{et~al.}~and Vashishth~\textit{et~al.}~\cite{rel_semisup_ea,rel_semisup_deep} take two different approaches to augment the input of deep learning models in semi-supervised learning situations. Jat~\textit{et~al.}~\cite{rel_semisup_ea} create two models: an Entity Attention (EA) model and a variation of a Recurrent Neural Network (RNN) model. They also consider an ensemble model comprised of these two new models and the Piecewise Convolutional Neural Network (PCNN)~\cite{rel_semisup_pcnn} which is successful for relation extraction. The word attention layer helps the EA model learn to increase the weight of important words, which assists in improving the prediction accuracy of the overall relation. This model has a better precision-recall curve than only the PCNN model~\cite{rel_semisup_pcnn}. Vashishth~\textit{et~al.}~\cite{rel_semisup_deep} leverage more information from KBs. They find ``side information" such as entity type and relation aliases. This side information is incorporated into the input of their Graph Convolutional Network (GCN) model~\cite{rel_semisup_gcn}. This model outperforms the entity/word attention model~\cite{rel_semisup_ea} and the PCNN model~\cite{rel_semisup_pcnn} showing the usefulness of additional information from KBs. 

\begin{table}[t]
\scriptsize
\begin{center}
\begin{threeparttable}
\caption{Comparison between precision, recall, and F-score of the state-of-the-art methods which are tested on the ChemProt~\protect\cite{rel_data_chemprot} dataset.}
\centering
\def\arraystretch{1.3}\tabcolsep 2pt
\label{table:ChemProt}
\begin{tabular}{|M{20.5mm}|M{20.5mm}|M{20.5mm}|M{20.5mm}|}
\hline
\textbf{Work} & \textbf{Precision (\%)} & \textbf{Recall (\%)} & \textbf{F-score (\%)}\\ 
\hline
\cite{lung2017extracting} & 62.52 & 51.21 & 56.71\\
\hline
\cite{matos2017extracting} & 57.38 & 47.22 & 51.81\\
\hline
\cite{lim2018chemical} & 67.04 & 51.94 & 58.53\\
\hline
\cite{peng2018chemical} & 72.66 & 57.35 & 64.10\\
\hline
\cite{verga2018simultaneously} & 48.00 & 54.10 & 50.80\\
\hline
\cite{liu2018extracting} & 57.40 & 48.70 & 52.70\\
\hline
\cite{corbett2018improving} & 62.97 & 62.20 & 62.58\\
\hline
\cite{mehryary2018potent} & 59.05 & 67.76 & 63.10\\
\hline
\cite{rel_sup_deep_gcn_protein} & 63.79 & 66.62 & 65.17\\
\hline
\end{tabular}
\end{threeparttable}
\end{center}
\end{table}

\begin{figure}[t]
\begin{center}
\includegraphics[width=\columnwidth]{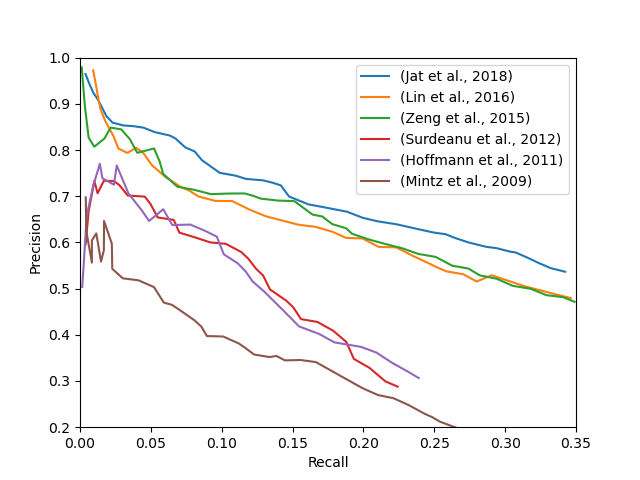}
\caption{Comparison between precision-recall curves of the state-of-the-art methods which are tested on the Riedel~2010~\protect\cite{rel_data_riedel} dataset. The compared works are: \protect\cite{rel_semisup_ea, lin2016neural, rel_semisup_pcnn, surdeanu2012multi, hoffmann2011knowledge, mintz2009distant}.}
\label{fig:Riedel_2010}
\end{center}
\end{figure}

A summary of relation extraction works can be found in Table~\ref{table:relation_extraction}. There are two common datasets for the recent works in Table~\ref{table:relation_extraction}, i.e., ChemProt~\cite{rel_data_chemprot}, and Riedel~2010~\cite{rel_data_riedel}. In order to fairly evaluate the results of the works, they have been compared with the ones which have used the same dataset. Table~\ref{table:ChemProt} lists the works that are tested on the ChemProt~\cite{rel_data_chemprot} dataset. \cite{rel_sup_deep_gcn_protein} has the highest F-score while \cite{peng2018chemical}, and \cite{mehryary2018potent} have the highest precision and recall, respectively. Furthermore, as it is depicted in Figure~\ref{fig:Riedel_2010}, \cite{rel_semisup_ea} has the best precision over the entire range of recall between the works that are compared on Riedel~2010~\cite{rel_data_riedel} dataset.

\subsection{Future Works \& Challenges}
Extracting relations of the form $t=(e_1,r,e_2)$ has been well-studied since the onset of the relation extraction task. However, recently, researchers have begun to look at alternative output forms to allow extracting more complex relations~\cite{rel_complex_relations}. McDonald~\textit{et~al.}~\cite{rel_orig_complex} explore complex relations, but they are limited by the fact that the structure of the relations has to be known prior to extraction. They note that altering the constraints in their problem setup could alleviate this limitation, but this suggestion has not been tested yet. Recent methods have started to look at this problem again and all apply Long-Short Term Memory (LSTM) methods~\cite{rel_n_ary,rel_n_ary_2,rel_n_ary_3}. Jia~\textit{et~al.}~\cite{rel_n_ary_2} also address document-level relation extraction, which is another limitation of relation extraction techniques. Most works only consider relations contained in one sentence as opposed to extracting relations that exist across a document.

As previously mentioned, Madaan~\textit{et~al.}~\cite{rel_semisup_minsup} study the problem of numerical relation extraction such that one entity in the extracted relation tuple is numeric. They note that their limitation is extracting the time scope of the relation. This is important because numerical relations, e.g. population, income, GDP, etc., often are only true for a certain time period, so the extracted relation is only valid for that period. To the best of our knowledge, no current works in relation extraction address numerical relation extraction or its associated time scope issue.

One of the biggest challenge in relation extraction tasks across all models is relation discovery which is a model's ability to identify a new relation that has never been seen before. This is still a limitation in semi-supervised models because such models tend to have many false-positives when generating patterns from seed facts~\cite{rel_semi_review}. This means that relation discovery in semi-supervised models is often noisy and dependent upon the seed facts provided.

\section{Open Information Extraction And Canonicalization}
\label{sec:open_information_extraction}
Similar to relation extraction, open information extraction refers to the extraction of relation tuples, $t~=~(a_1,r,a_2)$, from a text where $a_1,a_2$ are argument phrases and $r$ denotes the semantic relation between them~\cite{gamallo2014overview}. In other words, the main goal of OIE is to break syntactically complex sentences into relation tuples, which is used for various other tasks. The main difference of OIE compared to relation extraction is that the structures and types of relations do not need to be known in advance. This means that OIE models are better suited for relation discovery such that they can extract many new relations across domains. As a result, they scale to a large, heterogeneous collection such as the Web. Moreover, OIE avoids hand-labeling sentences and performs a single pass over a corpus with no pre-specified relations. Table~\ref{table:trad_oie_comparison} compares the relation and open information extraction key features.

\begin {table}[t]
\scriptsize
\begin{center}
\begin{threeparttable}
\caption{Relation and open information extraction comparison.}
\centering
\def\arraystretch{1.5}\tabcolsep 2pt
\label{table:trad_oie_comparison}
\begin{tabular}{|M{20mm}|M{31mm}|M{31mm}|}
\cline{2-3}
\multicolumn{1}{c|}{} & \textbf{Relation Extraction} & \textbf{OIE} \\ 
\hline
\textbf{Input} & Corpus + Hand-labeled data & corpus \\ \hline
Relations  & Specified in advance & Discovered automatically \\ \hline
Extractor & relation specific & relation independent \\ \hline
\end{tabular}
\end{threeparttable}
\end{center}
\end{table}

\subsection{General Terminology}

\subsubsection{Model Inputs}
There are common ways to describe the inputs of different models in OIE. The entities of interest are defined as an $n$-ary where $n$ represents the number of such entities that exist in an input sentence. In general, an input sentence is preprocessed with sequence labeling~\cite{li2020exploiting}. Sequence labeling is a Natural Language Processing (NLP) task where each word, also called a ``token", in a sentence is assigned a label. These model inputs are sometimes high dimensional vector spaces, so an initial step is translating such inputs to learned embeddings~\cite{zamani2017relevance} or low-dimensional vector representations.

Some methods leverage side information in their models. Side information~\cite{chechik2002extracting} is data that is not directly given as the input, but is useful to learn from during training. One example of side information is a knowledge base~\cite{murdaca2018knowledge}, an organized collection of facts about some domains. 

\subsubsection{Model Design Choices}
Sequence-to-Sequence (Seq2Seq)~\cite{sutskever2014sequence} is a deep learning-based model that takes a sequence of inputs e.g. words, letters, time series, etc. and outputs another sequence of the same type. A Long-Short Term Memory (LSTM)~\cite{hochreiter1997long} layer is typically used in Seq2Seq models. LSTMs contain feedback loops to retain information for a certain time period in order to process sequential data.

Another sequential model is Markov Decision Process (MDP)~\cite{white1989markov}. It provides a mathematical framework for modeling decision-making problems where outcomes of each decision happen with some probability such that they are partially random and partially determined by the previous decision. This model only considers sequences in one direction, while a BiLSTM model considers bidirectional sequences~\cite{liu2019bidirectional}. The BiLSTM model is composed of two LSTMs. One LSTM makes a forward pass through the input while the other makes a backward pass. In this way, the BiLSTM model learns the context of a given sentence as opposed to recognizing the individual words.

The other decision-making approach is Monte-Carlo Tree Search (MCTS) which is also considered as a sequential model. MCTS incrementally builds a tree. At each iteration, a node is selected and either a child node is added to it or the statistics of its ancestors are updated to balance between areas that have not been well sampled and areas that are promising~\cite{browne2012survey}.

\begin{figure}[t]
\begin{center}
\includegraphics[width=\columnwidth]{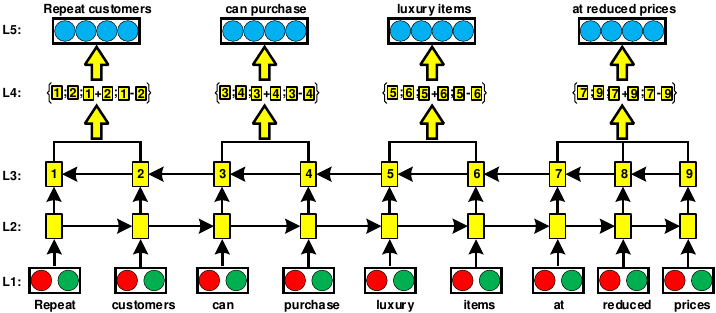}
\caption{The span model architecture~\protect\cite{Zhan2020SpanMF}. L1: An embedding is calculated for each word in the input sentence. L2 and L3: These embeddings are passed through a BiLSTM. L4: Several spans are produced. L5: These spans are pruned and scored so that the highest scoring spans are returned as the final output.}
\label{fig:oie_span}
\end{center}
\end{figure}

\subsection{Open Information Extraction}
Open information extraction systems extract tuples from input texts which represent relations conveyed by the inputs. These systems are used for tasks such as question answering~\cite{fader2014} and knowledge base population~\cite{angeli2015leveraging}. Most of OIE systems leverage semi-supervised approaches~\cite{etzioni2008open,wu2010open} or rule-based algorithms, the algorithms that match manually defined patterns~\cite{fader2011identifying,del2013clausie}, as they require less amount of work for gathering the labeled data. However, recent studies show that supervised learning can significantly improve the performance of OIE systems. Stanovsky~\textit{et~al.}~\cite{stanovsky2018supervised} propose a supervised approach where the information extraction is formulated as a sequence labeling problem. They extract multiple overlapping tuples for each sentence as well as calculate the extraction confidence to balance precision and recall. Beckerman~\textit{et~al.}~\cite{beckerman2019learning} propose implicit information extraction such that the extracted relation is not part of the input sentence. They only look at the subject and object of a sentence and infer the relation between them. This is different from other OIE methods where the verb in the sentence is typically considered part of the relation between the corresponding subject and object. To perform the implicit tuple extraction, they use a Seq2Seq model with attention layers.

Zhan~\textit{et~al.}~\cite{Zhan2020SpanMF} propose SpanOIE to address one of the main OIE challenges: training and testing models on an automatically built corpus that could generate meaningless sentences. This process may cause errors in the prediction as meaningless sentences train the model to recognize relations that do not exist in the real world. SpanOIE attempts to alleviate this drawback by leveraging a span model~\cite{ouchi2018span} which extracts predicate and argument ``spans" for each sentence. A span in this context means a subset of words in the given sentence. As it is depicted in Figure~\ref{fig:oie_span}, the span model generates multiple predicate spans with corresponding argument spans for a sentence. An inference score is used to order the generated spans and keep the highest scoring ones. In this way, meaningless sentences are reduced in the training set as the inference stage of the span model drops them. For testing, they solve the problem of meaningless test labels by manually re-annotating a widely used test set~\cite{spanoie}.

Another supervised OIE method is Monte-Carlo Tree Search (MCTS) based approach. Liu~\textit{et~al.}~\cite{liu2020extracting} use MCTS to overcome restricted search space limitations, whereas other models do not and potentially miss the optimal output due to lack of exploration. Their algorithm effectively increases the amount of tuples searched in order to find the optimal output tuple for a given input sentence. The first step in their algorithm is using a Seq2Seq model to generate phrases from the input sentence that could potentially be part of final extracted tuples. Next, they use MCTS to predict the likeliest sequence of generated phrases and, finally, these sequences are the output of the algorithm in the form of tuples. The larger search space used in this algorithm dramatically increases the search time to extract the optimal outputs. To address this, the authors propose a heuristic approach called Predictor Upper Confidence Bound (PUCB). PUCB estimates the overall likelihood of a phrase or delimiter occurring next in the sequence by looking at factors such as prior probability output by the Seq2Seq model and the number of times that phrase or delimiter has been searched. Further, PUCB takes into account the ``action value" of each phrase or delimiter, which is an estimation of its impact on predicting the rest of the sequence. Using these estimations, PUCB effectively eliminates phrases from the search space that do not have a high overall likelihood and decreases the search time.

\begin{figure}[t]
\begin{center}
\includegraphics[width=\columnwidth]{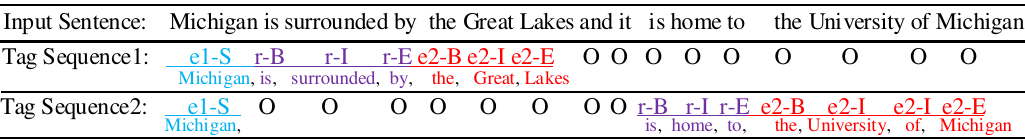}
\caption{An example of a sequence labeling scheme. Each tuple is independently labeled and multiple labels can be assigned for each word. Here, ``e1”, ``r”, ``e2” respectively denote entity1, relation, and entity2. Moreover, the BIOES annotation (``B": Begin, ``I": Inside, ``O": Outside, ``E": End, ``S": Single) is used for indicating the position of the token in an entity or the relational~\protect\cite{jia2019hybrid}.}
\label{fig:multiple_tuple_tagging}
\end{center}
\end{figure}

Advanced deep learning-based methods have become popular and achieved good accomplishments in various information extraction tasks. Jia~\textit{et~al.}~\cite{jia2019hybrid} leverage sequence labeling as a preprocessing stage where each tuple $t=(e_1,r,e_2)$ is independently labeled. This means that entities that exist in multiple tuples in the same sentence may be given different labels depending on their POS with respect to the relation of the tuple. An example of this preprocessing is shown in Figure~\ref{fig:multiple_tuple_tagging}. After this preprocessing stage, a Hybrid Neural Network (HNN)~\cite{chen2008hybrid} is used which leverages two kinds of neural networks simultaneously. The first is a surface learning agent, an RNN, that recognizes the local structure of subsets of the labeled sequence, while the other is a deep learning agent, a CNN, that learns its global structure. This means the surface agent looks for patterns in each individual token contained in the sequence and the deep agent learns patterns in the overall sequence. The use of these two agents leads to a better extraction accuracy. As a drawback of this approach, it does not consider n-ary relation tuples as opposed to~\cite{Zhan2020SpanMF} which limits the complexity of the relations that can be extracted.

\begin {table*}[t]
\scriptsize
\begin{center}
\begin{threeparttable}
\caption{Open information extraction and canonicalization methods summary. \tnote{$\dagger$}}
\centering
\def\arraystretch{1.3}\tabcolsep 2pt
\label{table:oie}
\begin{tabular}{|M{20mm}|M{25mm}|M{53mm}|M{40mm}|M{35mm}|}
\hline
\textbf{Work} & \textbf{OIE/Canonicalization} & \textbf{Main Technique} & \textbf{Advantage/Scope} & \textbf{Dataset} \\ 
\hline
TextRunner~\cite{etzioni2008open} & OIE & CRF & Web crawling to extract relations & Web documents from Yahoo\\ 
\hline
ReVerb~\cite{fader2011identifying} & OIE & Rule-based noun and relation phrase matching & Higher precision-recall area under the curve than~\cite{etzioni2008open} & - \\ 
\hline
OLLIE~\cite{schmitz2012open} & OIE & Pattern matching of dependency parse trees & Higher precision than~\cite{fader2011identifying} & - \\ 
\hline
ClausIE~\cite{del2013clausie} & OIE & Generate propositions from dependency parse tree of clauses & Higher precision than ~\cite{etzioni2008open,fader2011identifying,schmitz2012open} & Wikipedia, NYT~\cite{sandhaus2008new}\\
\hline
PropS~\cite{stanovsky2016getting} & OIE & Converts input to structured proposition format & Good format for QA tasks  & QA-SRL~\cite{fitzgerald2018large}\\ 
\hline
Jia~\textit{et~al.}~\cite{jia2019hybrid} & OIE & HNN & Multiple tokens per entity, simple to construct, and large-scale corpora & Wikipedia, NYT~\cite{sandhaus2008new}, Reverb~\cite{fader2011identifying}\\
\hline
RnnOIE~\cite{stanovsky2018supervised} & OIE & BiLSTM and BIO tagging: Extract multiple, overlapping tuples; Calculate extraction confidence &  Higher precision-recall area under the curve than~\cite{mesquita2013effectiveness, xu2013open} & AW-OIE, OIE2016, WEB-500, NYT-222, PENN-100~\cite{schneider2017analysing}\\
\hline
\cite{beckerman2019learning} & OIE & Implicit OIE & Implicit tuple: Maximally-shortened tuples with explicit common sense relations &  NewsQA~\cite{trischler2016newsqa}, SQuAD~\cite{rajpurkar2016squad}\\
\hline
SpanOIE~\cite{Zhan2020SpanMF} & OIE & Span model for n-ary OIE & & OIE2016~\cite{stanovsky2016creating}\\
\hline
MCTS~\cite{liu2020extracting} & OIE & MCTS + Seq2Seq model & Higher accuracy as a result of a larger search space & SAOKE~\cite{SAOKE}\\
\hline
SIST~\cite{lin2019canonicalization} & Canonicalization & Using side information from source text & Low time complexity & ReVerb45K~\cite{vashishth2018cesi} \\ 
\hline
FAC~\cite{wu2018towards} & Canonicalization & Pruning techniques: avoiding unnecessary computations, Bounding techniques: identifying small similarities & Fast speed& ClueWeb09~\cite{gabrilovich2013facc1}\\ 
\hline
CESI~\cite{vashishth2018cesi} & Canonicalization & Using embeddings \& side information & High accuracy & ReVerb45K~\cite{vashishth2018cesi}\\ 
\hline
\end{tabular}
\begin{tablenotes}\footnotesize
\item[$\dagger$] Abbreviation list: OIE: Open Information Extraction, CRF: Conditional Random Field, QA: Question Answering, HNN: Hybrid Neural Network, BIO: Begin - Inside - Outside, BiLSTM: Bidirectional Long-Short Term Memory, MCTS: Monte-Carlo Tree Search, Seq2Seq: Sequence-to-Sequence.
\end{tablenotes}
\end{threeparttable}
\end{center}
\end{table*}

\begin{table}[t]
\scriptsize
\begin{center}
\begin{threeparttable}
\caption{Comparison between AUC (Area Under
the precision-recall Curve) and F1-score for the state-of-the-art OIE systems which are tested on the OIE2016~\protect\cite{stanovsky2016creating} dataset.}
\centering
\def\arraystretch{1.3}\tabcolsep 2pt
\label{table:oie2016}
\begin{tabular}{|M{27.5mm}|M{27.5mm}|M{27.5mm}|}
\hline
\textbf{{Work}} & \textbf{{AUC}} & \textbf{{F1-score (\%)}}\\ 
\hline
OLLIE~\cite{schmitz2012open} & 0.202 & 38.58\\
\hline
ClausIE~\cite{del2013clausie} & 0.364 & 58.01\\
\hline
\cite{angeli2015leveraging} & 0.079 & 13.55\\
\hline
PropS~\cite{stanovsky2016creating} & 0.320 & 54.38\\
\hline
OpenIE4~\cite{mausam2016open} & 0.408 & 58.83\\
\hline
RnnOIE~\cite{stanovsky2018supervised} & 0.48 & 62\\
\hline
SpanOIE~\cite{Zhan2020SpanMF} & 0.489 & 68.65\\
\hline
\end{tabular}
\end{threeparttable}
\end{center}
\end{table}

Table~\ref{table:oie} summarizes the datasets, main techniques and pros and cons of each OIE approach. Moreover, Table~\ref{table:oie2016} compares the results of OIE systems which are tested on the OIE2016~\cite{stanovsky2016creating} dataset. SpanOIE~\cite{Zhan2020SpanMF} achieves the highest AUC (Area Under the precision-recall Curve) and F1-score.

\subsection{Canonicalizing Open Knowledge Base}
Recent advances in OIE approaches such as Nell~\cite{carlson2010nell} and ReVerb~\cite{fader2011identifying} have led to 
the creation and use of large structured knowledge bases~\cite{dong2014knowledge,bollacker2008freebase}. The entities and relations in these KBs are not canonicalized, which leads to storing redundant and ambiguous facts. Canonicalization, in this context, means storing information in KBs in a standardized format.

Generally, KBs contain a large amount of useful facts which are usually stored as $t=$(subject, relation, object) tuples. Without a canonicalization process, tuples such as (Barack Obama, was born in, Honolulu) and (President Obama, is president of, The United States) can both be stored as different entries in the same KB. However, since the subject part of each tuple is stored differently, both facts will not be retrieved when the entity ``Barack Obama" is queried~\cite{vashishth2018cesi}. So, it is impossible to guarantee that all facts about an entity will be retrieved. Several studies attempt to address this limitation of KBs~\cite{galarraga2014canonicalizing,lin2019canonicalization,wu2018towards,vashishth2018cesi}.

Gal{\'a}rraga~\textit{et~al.}~\cite{galarraga2014canonicalizing} canonicalize the KB in two steps. First, they partition the tuples into small groups, called canopies, based on which tuples share a common word. Second, they perform Hierarchical Agglomerative Clustering (HAC)~\cite{murtagh2014ward} on each canopy by merging tuples within it that have a high similarity score. If HAC is used without the canopies, each tuple must be compared to all others in the dataset. In this regard, their canopy method performs a smaller number of comparisons resulting in a lower time complexity than if canopies are not used. However, there may be a large number of canopies created with many overlapping tuples such that the lowered time complexity is still not efficient. For this purpose, Lin~\textit{et~al.}~and Wu~\textit{et~al.}~\cite{lin2019canonicalization,wu2018towards} propose efficient variants of~\cite{galarraga2014canonicalizing}.

Lin~\textit{et~al.}~\cite{lin2019canonicalization} improve upon the HAC algorithm by making the size of canopies larger with, including tuples that may not have a common word between all of them, but they share a common word with some tuples in the canopy. This means that the number of canopies is reduced and there is less tuple overlap between canopies. The time complexity of canonicalizing the KB is smaller than~\cite{galarraga2014canonicalizing}. On the other hand, Wu~\textit{et~al.}~\cite{wu2018towards} improve efficiency by pruning large canopies through various criteria and bounding techniques. These bounding techniques include computing similarity scores between entities only when attaching an edge between them in their canopy and an indexing process that maps a set of words to each tuple that these words are mentioned in. Both of these techniques help in identifying tuples that exist in the same canopy without needing to compute extra similarity scores for tuples that do not exist in the same canopy.

\begin{figure}[t]
\includegraphics[width=\columnwidth]{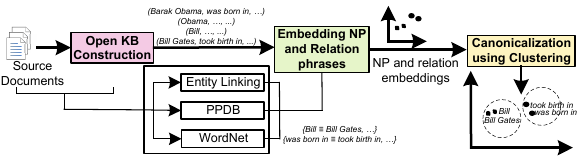}
\caption{An overview of CESI. CESI after gathering side information from noun and relation phrases of open KB triples, learns embedding of these noun phrases using the side information. Then, it performs clustering over the learned embeddings~\protect\cite{vashishth2018cesi}.}
\label{fig:cesi}
\end{figure}

Vashishth~\textit{et~al.}~\cite{vashishth2018cesi} propose Canonicalization using Embeddings and Side Information (CESI) which also leverages HAC. However, as opposed to~\cite{galarraga2014canonicalizing,lin2019canonicalization,wu2018towards}, CESI first preprocesses the tuples into a learned embedding to make the input lower-dimensional before using HAC. To learn this embedding, CESI minimizes an objective function consisting of the sum of HolE's~\cite{nickel2016holographic} objective function and terms related to side information. HolE's objective function is chosen because of its high performance on embeddings used for knowledge graphs. The side information terms in the objective function include entity linking~\cite{spitkovsky2012cross}, PPDB information~\cite{pavlick2015ppdb}, WordNet with word-sense disambiguation~\cite{banerjee2002adapted,miller1995wordnet}, etc. CESI then performs HAC on those embeddings and outputs canonicalized noun and relation tuples which can now be added to a KB. The architecture of CESI is shown in Figure \ref{fig:cesi}.

\begin {table}[t]
\scriptsize
\begin{center}
\begin{threeparttable}
\caption{Comparison between relation canonicalization results of the state-of-the-art methods which are tested on the ReVerb45K~\protect\cite{vashishth2018cesi} dataset. ReVerb45K~\protect\cite{vashishth2018cesi} includes 22K relation phrases.}
\centering
\def\arraystretch{1.3}\tabcolsep 2pt
\label{table:ReVerb45K}
\begin{tabular}{|M{16mm}|M{14.8mm}|M{14.8mm}|M{14.8mm}|M{20mm}|}
\hline
\textbf{Work} & \textbf{Macro Precision} & \textbf{Micro Precision} & \textbf{Pairwise Precision} & \textbf{Induced Relation Extraction}\\ 
\hline
PATTY~\cite{nakashole2012patty} & 78.2\% & 87.2\% & 80.2\% & 1814\\
\hline
AMIE~\cite{galarraga2013amie} & 69.3\% & 84.2\% & 66.2\% & 51\\
\hline
CESI~\cite{vashishth2018cesi} & 77.3\% & 87.8\% & 72.6\% & 2116\\
\hline
SIST~\cite{lin2019canonicalization} & 87.5\% & 87.2\% & 84.5\% & 1701\\
\hline
\end{tabular}
\end{threeparttable}
\end{center}
\end{table}

Table~\ref{table:ReVerb45K} compares the canonicalization methods that are tested on the ReVerb45K~\cite{vashishth2018cesi} dataset. SIST~\cite{lin2019canonicalization} has the highest macro precision while CESI~\cite{vashishth2018cesi} achieves the highest micro precision and induced relation extraction.

\subsection{Future Works \& Challenges}
A key step in OIE is the evaluation of extracted tuples, and this area needs further development because a clear and formal specification of what a valid relational tuple consists of is still missing. Most OIE systems evaluate their approaches based on a wide range of datasets, which makes validating and comparison of them more complicated. Most studies construct their own benchmark to evaluate their achievements. Thus, a benchmark dataset or a standardized evaluation method must be constructed for the OIE task for a valid comparison between works.

Another important aspect of OIE is how well OIE systems scale to larger databases. Current methods are inefficient for larger scales, so they cannot efficiently extract information from many databases~\cite{adnan2019analytical}. In order for OIE techniques to be used more effectively, OIE systems should be able to handle larger databases of unstructured text. 

Finally, a potential improvement in OIE systems could be in the canonicalization of their outputs. This ties in with the standardization of OIE evaluation as the outputs of OIE systems should be in similar formats so that they may be easily evaluated using a standard evaluation technique or benchmark dataset. This would allow for efficient advancements in OIE as it will be clear when one system has an advantage over another. 

\section{Open Information Extraction Applications}
\label{sec:application}
OIE’s flexibility from not being restricted to a fixed relation vocabulary makes it robust and powerful to be applied on many different types of textual resources~\cite{mausam2016open}. Considering the various strengths of OIE, this section describes semantic tasks that can incorporate OIE.

\subsection{General Terminology}
\subsubsection{Semantic Tasks}

Semantics is the meaning of a language in a text that goes beyond the grammatical structure. Semantics are necessary to provide additional information which helps in understanding textual data~\cite{mihalcea2001word}. Different semantic applications, such as text summarization and knowledge graph construction, require extracting useful textual units from an intermediate data representation structure that has been constructed in the preprocessing stage~\cite{stanovsky2015openintermediate}. Lexical representations, dependency parse trees, and Semantic Role Labeling (SRL) are common intermediate structures being used in the preprocessing stage.

In general, the lexical representations such as lexical chains focus on the original word sets instead of building structures or connections on the input documents. It captures the features such as lexical overlap or cohesion among an arbitrary number of related words directly from the input texts~\cite{silber2002efficiently}. Dependency parse trees introduce a tree structure to represent the relations between words within a sentence and construct non-local dependencies such as subject and auxiliary around a single predicate. Figure~\ref{fig:rel_sup_tree} is an example of a dependency parse tree. It is helpful to analyze long and complex sentences~\cite{cer2010parsing, fundel2007relex}. SRL specifies a list of potential semantic roles for a single word or phrase in advance, such as a label for the agent performing an action or for the goal of the action in a sentence. It then extracts the semantic relations drawn from that list between the predicate and its corresponding participants and properties~\cite{marquez2008semantic}. Unlike the aforementioned intermediate structures, OIE extracts the propositions that contain multi-word predicates and infinitive phrases in a single predicate slot. For example, OIE will extract ``agree to pay" and ``take care of" from the sentence ``Mary agrees to pay a large amount of extra fees to let Harry take care of her child" instead of separating the predicates into single words (e.g. ``agree", ``to", and ``pay"). Moreover, the arguments in the structure extracted by OIE are not analyzed to certain semantic roles as SRL does and are truncated in specific cases such as prepositional phrases. This difference makes OIE more human-readable. Also, within a single representation, OIE can recognize related words, even when they are not in close proximity in the representation as well as identify words that have similar meanings.

It has been shown that OIE positively impacts the performance of semantic tasks that depend on such an intermediate step~\cite{mausam2016open, stanovsky2015openintermediate}. Thus, the rising trend of OIE’s development provides a new research direction of applying it to the semantic tasks.

\subsubsection{Resource Description Framework}
\begin{figure}[t]
\includegraphics[width=\columnwidth]{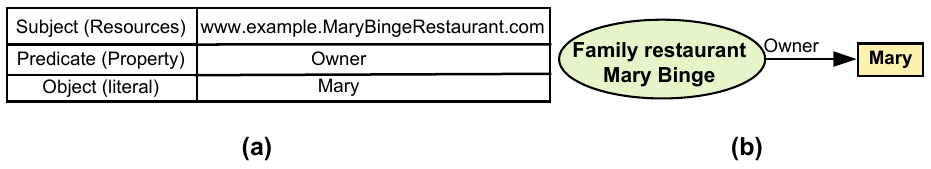}
\caption{An example of a basic RDF for \emph{``Mary is the owner of the family restaurant Mary Binge as shown on www.example.MaryBingeRestaurant.com"} sentence. a) Demonstrates three object types: resources, properties, and statements. b) Illustrates a pictorial representation for this basic RDF model example~\protect\cite{lassila1998resource}.}
\label{fig:rdf}
\end{figure}

Resource Description Framework (RDF) is the foundation for processing metadata through the Web and stands for a standard model of metadata to interchange machine-understandable data on the Web~\cite{lassila1998resource}. It can be considered as a simple framework that represents expressions in an easily applicable way. There are three main object types in RDF expressions. Resources are sources (e.g. web pages or a set of online documents) that the RDF describes, properties are attributes of the resource, and statements are the combination of a resource, a property of that resource, and the value of that property. Figure~\ref{fig:rdf} shows an example of RDF.

\subsection{Knowledge Graph}
\begin{figure}[t]
\includegraphics[width=\columnwidth]{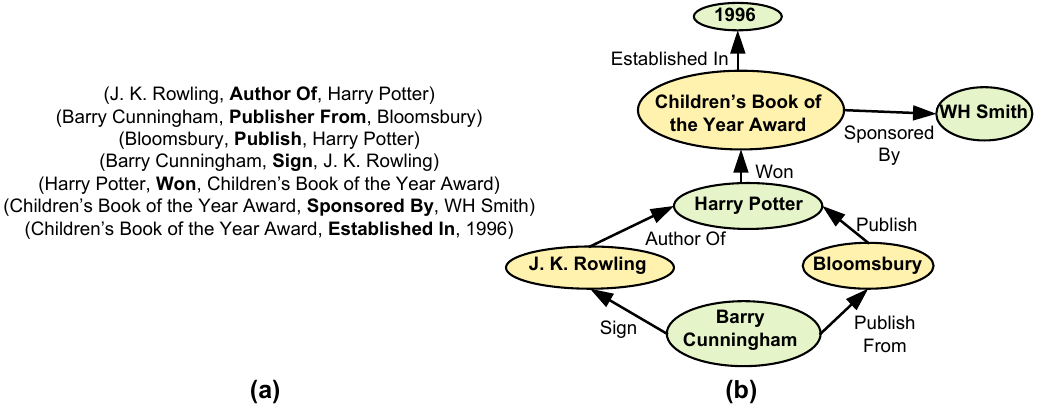}
\centering
\caption{An example of knowledge graph representations. a) Factual triple under the Resource Description Framework (RDF). b) Directed graph with nodes as entities and edges as relations~\protect\cite{ji2020survey}.}
\label{fig:knowledgegraph}
\end{figure}

Knowledge Graph (KG) is a structured way of representing facts, consisting of entities, relationships between them, and semantic descriptions of both entities and relations. This structure allows a KG to convey meaningful connections between facts such that knowledge semantics, rich ontologies, and multi-lingual knowledge can be represented~\cite{ji2020survey}. Figure~\ref{fig:knowledgegraph} shows an example of how a KG represents the information. Representing unstructured text in a structured KG is powerful and effective for further usages such as text generation and search engine suggestions~\cite{ji2020survey}.

There are several issues regarding constructing KGs using OIE. First, there is entity selection, meaning picking only one entity when multiple entities appear as the subject or object of the relation, and, second, the fact that representation is limited to one tuple. On the other hand, OIE semi-structured outputs of the form $t=$ (\textit{subject, relation, object}) allows for easy conversion into RDF triples to be used for a KG as no training or complex rules are necessary for this conversion. Martinez-Rodriguez~\textit{et~al.}~\cite{martinez2018openie} propose a method to generate KGs using relations extracted from OIE methods. They convert OIE outputs to RDF triples in the following steps. Extracting the entities in the subject and object as well as their relation, determining the tense and conjugation of the relation, and, finally, combining all of this information into an RDF triple.

OIE is one of the best performing methods to extract SVOs for KGs~\cite{chu2017distilling}. Dess{\`\i}~\textit{et~al.}~\cite{dessi2020ai} propose Artificial Intelligence Knowledge Graph (AI-KG), which is a large-scale automatically generated knowledge graph that provides a description of research publications in the AI field. They apply OIE to parse the text and build a set of tuples of the form $t=$ (\textit{Subject, Verb, Object}) (SVO). A subset of these relations with both subject and object that are overlapped with the research entities extracted by DyGIE++~\cite{wadden2019entity}. Finally, CSO classifier~\cite{salatino2019cso} forms the final set of valid relations. Li~\textit{et~al.}~\cite{li2018improving} construct a KG that describes caveats in Android Application Programming Interfaces (APIs). API caveats are instances where the API is used incorrectly, so such a knowledge graph is useful in helping programmers avoid mistakes. They leverage OIE as the first step to extract SVOs from sentences about API caveats which are gathered from existing API documentations. The subjects and objects of the tuples have a unique link to their original API documentation source. So, the relation between the source and the tuple is preserved in the KG. Moreover, Al-Khatib~\textit{et~al.}~\cite{al2020end} use the arguments and predicates of OIE and SRL outputs and to compose a set of syntactic patterns for the concept identification task, which is an important part of the knowledge required to construct KGs. The experiment results show that OIE patterns achieve better accuracy than the SRL patterns.

Apart from being used for a subprocess of constructing KGs, OIE is sometimes used to evaluate KGs. For instance, scientific KGs aim to describe complex dependencies of conditions in the scientific observation, and OIE is used as a baseline to verify the effectiveness of these KGs~\cite{jiang2019role}.

\subsection{Text Summarization}
Automatic text summarization is the task of producing a concise and smooth summary text that describes the key content information while preserving the overall meaning of the original texts~\cite{allahyari2017textsum9}. As an example, if we have the original text as: \emph{``The girl has been looking forward to going to see a new movie all week, and she has finished all her homework ahead of time so she is free to see the movie"}, then it can be reduced to be \emph{``The girl is excited to see the new movie."}. In this example, we can see the core meaning of the original text is preserved in the concise summarization. Text summarization nowadays has a broad application in all fields~\cite{colter2022tablext,fayazi2022fascinet}.

In general, a summary sentence should be informative and nonrepetitive. The semi-structured output of OIE methods is useful for this goal because it is possible to identify redundancy. Christensen~\textit{et~al.}~\cite{christensen2013towards10} use OIE to handle the redundancy using an unsupervised approach. Their method is able to identify inferred edges between related summary sentences. An edge is a link connecting two sentences and is directed so that it preserves the original ordering of the sentences to ensure the coherence of the summary texts. One edge can be inferred from the other if both edges have an overlapping component, e.g. sentence $b_1$ in $(a,b_1)$ and $b_2$ in $(a,b_2)$ are similar in meaning since $a$ is an overlapping component. To identify whether a pair of sentences contain similar information, OIE is used to extract the relational tuples.

OIE is also applied to train the vector embeddings for words in similarity and analogy tasks. These embeddings can be trained in such a way so that they output a similarity score indicating how similar two words are. Stanovsky~\textit{et~al.}~\cite{stanovsky2015openintermediate} evaluate the training of embeddings on a subset of Wikipedia pages with four different data representation structures i.e. lexical context, dependency parse trees, SRL’s semantic relations, and OIE’s tuples. The comparison results of the similarity scores computed from these embeddings show that OIE-based embeddings consistently outperform the others. Moreover, these OIE-based embeddings can better identify the similar word-pairs revealing topical similarity which implies that two words belong to similar topics (e.g. truck, transportation), and functional similarity which means that two words represent similar functions (e.g. truck, bus, car).

Other than utilizing OIE to only deal with the sentence or word similarity, Rahat~\textit{et~al.}~\cite{rahat2018open12} propose a novel text summarization technique that directly uses the propositions extracted by OIE to identify the most informative parts of each sentence. Those propositions will be merged in the later steps to form the summary text. 

\subsection{Question Answering Tasks}
OIE provides a way to generate knowledge for Question Answering (QA). QA is the task of retrieving answers from unstructured text through queries in question format~\cite{allam2012question}. QA can be performed in a closed-domain or open-domain format. Closed-domain QA is limited to answering queries within a specific field, while open-domain QA is applied on open-domain datasets~\cite{ramprasath2012survey}.

\begin{figure}[t]
\includegraphics[width=\columnwidth]{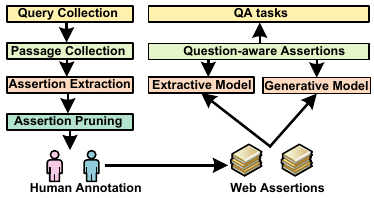}
\centering
\caption{Flow of ABQA dataset construction and how it can be leveraged in other question answering tasks~\protect\cite{yan2018assertion}.}
\label{fig:abqa}
\end{figure}

Several specific QA tasks have been proposed~\cite{voorhees1999trec,rajpurkar2016squad,yan2018assertion} with different answer formats and datasets. Voorhees~\textit{et~al.}~\cite{voorhees1999trec} define the answer format as a passage of text that contains the answer to the input question. Rajpurkar~\textit{et~al.}~\cite{rajpurkar2016squad} propose the Reading Comprehension (RC) task and construct a dataset of passages with corresponding questions and answers for each passage such that systems only select the correct answer out of a database of passages to the input question instead of returning a whole passage of text relevant to the input question. Yan~\textit{et~al.}~\cite{yan2018assertion} propose a new task, Assertion-Based Question Answering (ABQA), which is creating answers for input questions with a semi-structured assertion of the form $a=$~\textit{(subject,~predicate,~object)}. Figure~\ref{fig:abqa} gives an overview of how ABQA dataset is constructed as well as the application of ABQA in other QA tasks. This semi-structured assertion output improves upon the other tasks~\cite{voorhees1999trec,rajpurkar2016squad} such that this output is more concise than a passage but provides more information than a single answer output.

Some studies focus on answering simple questions~\cite{fader2014,yin2015answering}, while others attempt to tackle more challenging problems~\cite{yan2018assertion,Khot2017AnsweringCQ,khashabi2016question}. Khashabi~\textit{et~al.}~\cite{khashabi2016question} present a new Integer Linear Programming (ILP)-based model, TABLEILP, that answers questions using manually constructed tables as a KB. They use a support graph optimization framework~\cite{khashabi2016question} to create the best supported answer by text within available documents in a KB. A support graph connects words in the input question to words in pieces of text in the KB which helps formulate an answer. The support graph is optimized through ILP. Khot~\textit{et~al.}~\cite{Khot2017AnsweringCQ} propose TUPLEINF which uses the same framework, but leverages a potentially noisy KB that is constructed from OIE methods. TUPLEINF outperforms other QA systems such as~\cite{khot2015exploring} which uses a small set of logical rules and TABLEILP~\cite{khashabi2016question}. TUPLEINF addresses more diverse and complex questions while also eliminating the dependency on manually selected knowledge as it only leverages information automatically gathered from OIE methods.

Yan~\textit{et~al.}~\cite{yan2018assertion} propose two methods for the ABQA task. They create a generative model, Seq2Ast, based on the Seq2Seq training objective~\cite{sutskever2014sequence} and an extractive model, ExtAst, based on a ranking objective. Seq2Ast uses a bidirectional Gated Recurrent Unit (GRU) to encode the input. Then, they develop a hierarchical decoder~\cite{ronanki2017hierarchical} to decode the encoding and produce the output assertion. For Seq2Ast, the tuple structure of the assertion is first decoded and then the specific words in each part of the tuple are decoded. ExtAst selects an answer to the input question through three steps. The first step is generating possible output assertions with an OIE model, ClausIE~\cite{del2013clausie}. Next, features at the word, phrase, and sentence level are compared between each possible assertion and the question. Finally, a ranking algorithm, LambdaMART~\cite{burges2010ranknet}, is used to choose the best answer. 

\begin{table*}[t]
\scriptsize
\begin{center}
\begin{threeparttable}
\caption{OIE applications summary. \tnote{$\dagger$}}
\centering
\def\arraystretch{1.3}\tabcolsep 2pt
\label{table:OpenIE_application}
\begin{tabular}{|M{35mm}|M{16mm}|M{69mm}|M{55mm}|}
\hline
\textbf{Work} & \textbf{Application} & \textbf{Main technique} & \textbf{Advantage/Scope}\\ 
\hline
OIE-based KG construction~\cite{martinez2018openie} & KG & Extracting atomic units of relational information & Extract meaningful propositions from text\\ 
\hline
AI-KG~\cite{dessi2020ai} & KG & Extracting SVO triples as a subset to construct final set of relations & General performance\\ 
\hline
API caveats KG~\cite{li2018improving} & KG & Extracting SVO triples from API caveat sentences for open linking & General performance\\ 
\hline
TABLEILP~\cite{khashabi2016question} & QA & ILP-based model, support graph optimization framework & Manually constructed tables as a KB \\ \hline
TUPLEINF~\cite{Khot2017AnsweringCQ} & QA & ILP-based model, support graph optimization framework & Using OIE constructed KBs, Addressing diverse \& complex questions, Dealing with multiple short facts\\ 
\hline
Seq2Ast~\cite{yan2018assertion}, ExtAst~\cite{yan2018assertion} & QA & BiGRU-based RNN encoder and hierarchical decoder, LambdaMART~\cite{burges2010ranknet} ranking algorithm used to pick answer from OIE generated assertions & Generate an answer that is not directly asserted in the input text, Pick an answer that is clearly stated in the input text with a reliable OIE method\\ \hline
\end{tabular}
\begin{tablenotes}\footnotesize
\item[$\dagger$] Abbreviation list: KG: Knowledge Graph, SVO: Subject, Verb, Object, QA: Question Answering, ILP: Integer Linear Programming, KB: Knowledge Base, BIGRU: Bidirectional Gated Recurrent Unit, RNN: Recurrent Neural Network.
\end{tablenotes}
\end{threeparttable}
\end{center}
\end{table*}

\subsection{Future Works \& Challenges}
\subsubsection{Text Classification}
Classification is the task of categorizing information by analyzing its contents. Recently, many methods have been studied to achieve high accuracy and efficiency. Naive Bayes (NB) classifiers~\cite{1998naiveReview} and Support Vector Machines (SVM)~\cite{2001SVM} are the dominant approaches that have been leveraged for this purpose. However, most current classifiers with state-of-the-art accuracy rely on Deep Neural Networks (DNN)~\cite{ModernNNApplication0, ModernNNApplication1}.

A classification application is a topical classification, where one or more categories are selected from a larger set of categories. A common example of this is topical web search classification, where queries are classified to return relevant information~\cite{WebTopicalApplication}. There are many opportunities for applying OIE in the classification task. One such example is aiding accurate text extraction from PDFs~\cite{pdfClass}. PDFs are notoriously difficult for machines to read due to their close proximity of meta-data and text data. A topical classifier that takes advantage of OIE principles can accurately classify all of the text through the conversion of unstructured text to structured relational tuples as OIE provides.

In general, OIE works in the domain of unstructured text which is where most text classification, such as web query classification, occurs. So, the techniques used in OIE could be useful as a preprocessing stage to text classifiers such that classifier algorithms can be built using extracted relation tuples as inputs.

\subsubsection{Semantic-based Keyword Extraction}
Keyword extraction’s main task is to identify the important textual units within a document that can best express its content~\cite{turney2000learning, kelleher2005automatic}. A baseline algorithm for keyword extraction is the Kea algorithm~\cite{witten2005kea}. One of the main performance improvements from OIE in keyword extraction can be encoding semantic relations into the models to obtain more information of the candidate keyword is obtained. 

In the Kea algorithm, term frequency and the location of the first occurrence of a term are used to train a model that learns to extract key phrases from documents. Kea++~\cite{medelyan2008domain} is an extension of the Kea algorithm that includes a semantic feature, node degree of a phrase, that counts the number of candidate phrases in the thesaurus that are semantically linked to it. Li~\textit{et~al.}~\cite{li2014lexicalchain} add the lexical chain, a set of semantically related word sequences to capture the cohesive structure of a document, to Kea in order to improve output quality. The results show the main area of improvement in this algorithm is the way features are defined in the training stage. It is possible that the relational SVOs extracted by OIE can be incorporated in the training stage through certain computation methods as another way to reveal how important a subject or object is. The advantage of using the OIE outputs would be in either capturing relations between key phrases or extracting more complete key phrases.

\section{Conclusion}
\label{sec:conclusion}
This paper studies the recent open information extraction techniques and their applications. After reviewing the main concepts and approaches of relation extraction as the main OIE basis, recent OIE and canonicalization methods have been analyzed. Moreover, the advantages and disadvantages of each work are summarized and the current challenges in addition to the potential future works. Finally, the applications of OIE in knowledge graph, text summarization, question answering, text classification, and semantic-based keyword extraction have been discussed.

\ifCLASSOPTIONcaptionsoff
  \newpage
\fi
\bibliographystyle{IEEEtran}
\balance

\end{document}